%% file: main.tex
\documentclass[a4paper,10pt]{article}

\usepackage[utf8]{inputenc} 
\usepackage[a4paper,
            bindingoffset=0.2in,
            left=1in,
            right=1in,
            top=1in,
            bottom=1in,
            footskip=.25in]{geometry}
            
\input{packages.tex}

\usepackage[url=false, doi=true,backend=biber,maxbibnames=6,maxcitenames=20,giveninits=true,defernumbers=true,isbn=false,sorting=none]{biblatex}
\usepackage{biblatex-shortfields}
\makeatletter
\let\blx@rerun@biber\relax
\makeatother

\begin{document}
\begin{acronym}
	\acro{rhs}[RHS]{right-hand side}
	\acrodefplural{rhs}[RHSs]{right-hand sides}
	\acro{2d}[2D]{two-dimensional}
	\acro{3d}[3D]{three-dimensional}
	\acro{ac}[AC]{alternating current}
	\acro{am}[AVM]{adjoint variable method}
	\acro{dc}[DC]{direct current}
	\acro{dof}[DoF]{degrees of freedom}
	\acro{dsm}[DSM]{direct sensitivity method}
	\acro{epdm}[EPDM]{ethylene propylene diene monomer}
	\acro{eqs}[EQS]{electroquasistatic}
	\acro{eqst}[EQST]{electroquasistatic-thermal}
	\acro{es}[ES]{electrostatic}
	\acro{fe}[FE]{finite element}
	\acro{fem}[FEM]{Finite Element Method}
	\acrodefplural{fgm}[FGMs]{field grading materials}
	\acro{fgm}[FGM]{field grading material}
	\acro{hv}[HV]{high voltage}
	\acro{hvac}[HVAC]{high voltage alternating current}
	\acro{hvdc}[HVDC]{high voltage direct current}
	\acro{ode}[ODE]{ordinary differential equation}
	\acrodefplural{pde}[PDEs]{partial differential equations}
	\acro{pde}[PDE]{partial differential equation}	
	\acrodefplural{pec}[PECs]{perfect electric conductors}
	\acro{pec}[PEC]{perfect electric conductor}
	\acrodefplural{pec}[PECs]{perfect electric conductors}
	\acrodefplural{qoi}[QoIs]{quantities of interest}
	\acro{qoi}[QoI]{quantity of interest}
	\acro{rms}[rms]{root mean square}
	\acro{sir}[SiR]{silicone rubber}
	\acro{xlpe}[XLPE]{cross-linked polyethylene}
\end{acronym}
\pagenumbering{gobble}

\date{}
\title{\Large\bf 
Transient Nonlinear Electrothermal Adjoint Sensitivity Analysis for HVDC Cable Joints}
\author{
        $^{1,2}$M. Greta Ruppert, $^{1,2}$Yvonne Späck-Leigsnering,\\ $^{1,2}$Herbert De Gersem	\vspace{1ex}  \\
	{\it  $^1$ Technische Universität Darmstadt, Institute for Accelerator Science and}\\{\it Electromagnetic Fields, Schlossgartenstr.~8, 64289 Darmstadt, Germany}\\
	{\it  $^2$ Technische Universität Darmstadt, Graduate School of Computational Engineering }\\{\it  Dolivostraße~15, 64293 Darmstadt, Germany}
}

\maketitle
\let\thefootnote\relax\footnotetext{\textcolor{red}{Published in the International Journal of Numerical Modelling: Electronic Networks, Devices and Fields, 28th April 2025. Please cite using the DOI: https://doi.org/10.1002/jnm.70035}}

{\large \textbf{Abstract }} 
Efficient computation of sensitivities is a promising approach for efficiently designing and optimizing \acl{hvdc} cable joints. This paper presents the \acl{am} for coupled nonlinear transient electrothermal problems as an efficient approach to compute sensitivities with respect to a large number of design parameters. 
The method is used to compute material sensitivities of a 320\,kV \acl{hvdc} cable joint specimen. The results are validated against sensitivities obtained via the \acl{dsm}. 
\section{Introduction}	
	\input{introduction.tex}
\section{Modeling and Numerical Approach}
	\subsection{Cable Joint Specimen}
	\label{sec:specimen}
 \label{ch:joint_specimen}
 \begin{figure}[tb]%
	\centering
	\includegraphics[width=0.6\columnwidth]{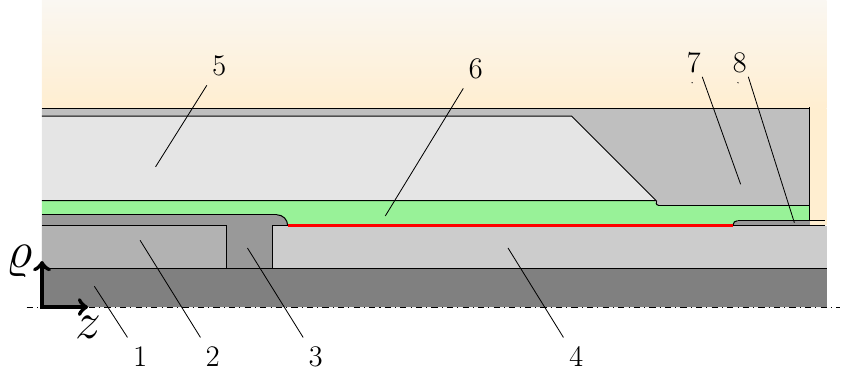}
\vspace{0.2cm}
\caption{\begingroup \small Schematic of the investigated \ac{hvdc} cable joint \cite{Hussain_2017aa} in a cylindrical coordinate system $(\varrho,z)$. The numbers indicate the different materials as described in the text, the \ac{fgm} layer is highlighted in green. The red line marks the interface between the \ac{fgm} layer and the cable insulation. The cable joint is surrounded by a 30\,cm thick layer of sand and buried 2\,m beneath the ground. 
\endgroup}\label{Muffe Konfiguration}
\end{figure}
\begin{figure}[tb]	
	\centering
	\setlength{\fwidth}{0.45 \columnwidth}
	\input{figs/ftsigma_RH.tex}
	\caption{Nonlinear field- and temperature-dependent conductivity of the \ac{fgm}. The \ac{fgm} conductivity is modeled by the analytic function \eqref{Conductivity}.}
	\label{fig:sigma_general}
\end{figure}
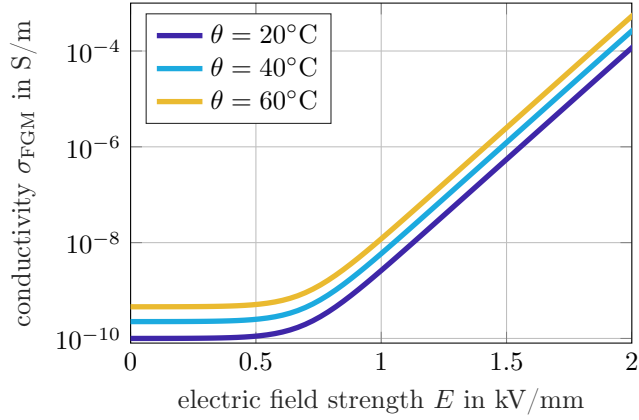

Figure~\ref{Muffe Konfiguration} shows a cut of the investigated 320\,kV \ac{hvdc} cable joint model \cite{Hussain_2017aa}. 
The cable joint connects two cables segments, each consisting of a copper conductor (domain 1), an insulation layer (domain 4) and a grounded outer sheath (domain 8). The conductors are connected by an aluminum connector (domain 2), which is encased in conductive \ac{sir} (domain 3). The primary insulation of the cable (domain 4) is composed of \ac{xlpe}, while the main insulation of the cable joint body (domain 5) is made from insulating \ac{sir}. The outer sheath of the cable joint (domain 7) as well as the outer semiconductor of the cable (domain 8) are grounded. 
The cable joint is surrounded by a 30\,cm thick layer of sand and buried 2\,m beneath the ground. 

In the simulation, a switching impulse overvoltage is applied to the domains 1 to 3, which are modeled as \acp{pec} due to their substantially higher conductivities relative to the insulating materials. 
The assumed operating temperatures are 65$^\circ$C at the copper conductors and 20$^\circ$C for the ambient temperature at the top of the soil layer \cite{Spack_2021ab}. For a detailed description of the geometric measurements of the cable joint and material characteristics, see \cite{Spack_2021ab, Hussain_2017aa}.

To mitigate harmful electric field stresses at the insulation interfaces between the \ac{xlpe} and \ac{sir}, a layer of nonlinear \ac{fgm} (domain 6) is introduced. This \ac{fgm} layer dynamically adjusts its conductivity based on the local electric field strength, $E$. In regions of high electric field stress, the \ac{fgm}'s conductivity increases, thereby redistributing the electric field by shifting the voltage drop to less stressed areas. This results in a more balanced electric field distribution across the insulation layers. 

The nonlinear conductivity of the \ac{fgm}, $\sigma_{\+{FGM}}$, is modeled by the following function \cite{Hussain_2017aa}:
\begin{equation}\label{Conductivity}
\begin{split}
\sigma_{\+{FGM}}&(E,\theta)=p_{1}\frac{1+p_{4}^{(E-p_{2})p_{2}^{-1}}}{1+p_{4}^{(E-p_{3})p_{2}^{-1}}}\+{exp}(-p_{5}(\theta^{-1}-\theta_{0}^{-1}))\,,
\end{split}
\end{equation}
where $\theta$ is the temperature, $\theta_0 = 293.15\,K$ is the reference ambient temperature, and $p_1$ to $p_5$ are design parameters. Selecting suitable values for $p_1$ to $p_5$ is a non-trivial task, as each parameter significantly influences the \ac{fgm}'s nonlinear field-grading behavior. For example, $p_1$ determines the baseline conductivity, while $p_2$ specifies the electric field strength where the \ac{fgm} transitions from baseline to nonlinear response. 
In this study, the parameters are initially set to $p_1 = 10^{-10}\,\text{S/m}$, $p_2 = 0.7\,$kV/m, $p_3 = 2.4$\,kV/m, $p_4 = 1864$ and $p_5 = 3713.59\,\+K$, with the field dependence at a fixed temperature illustrated in Fig.~\ref{fig:sigma_general}.

	\subsection{Electrothermal Modeling}
		\input{electrothermal_modeling.tex}
	\subsection{Discretization in Space and Time}
		\input{eqst_numerical_approach.tex}
\section{Sensitivities of Cable Joint Materials}
	\input{sensitivities_general.tex}
\subsubsection{Discretization in Space and Time}
	\input{avm_eqst_numerical_approach.tex}
\section{Simulation Studies}
	\input{results.tex}

\section{Conclusion}
	\input{conclusion.tex}
\section{Acknowledgements}
The authors thank Rashid Hussain for providing the simulation model and material characteristics published in \cite{Hussain_2017aa}. This work is supported by the DFG project 510839159, the joint DFG/FWF Collaborative Research Centre CREATOR (CRC–TRR361/F90) and the Graduate School Computational Engineering at the Technische Universität Darmstadt. Yvonne Späck-Leigsnering holds an Athene Young Investigator Fellowship of the Technische Universität Darmstadt. 

%

\printbibliography

\end{document}

%% file: packages.tex
\usepackage{amsmath}   
\usepackage{ amssymb }
\usepackage[utf8]{inputenc}
\usepackage[T1]{fontenc}
\usepackage{footnote}
\usepackage[english]{babel} 
\usepackage{hyperref}         
\usepackage{graphicx}         
\usepackage{epstopdf}	      
\usepackage{subfig}
\usepackage{bm} 				
\usepackage{booktabs}
\usepackage{color}
\usepackage{float}
\usepackage{upgreek}
\usepackage{ textcomp }
\usepackage{empheq}

\usepackage[nolist]{acronym}

\newcommand{\+}[1]{\text{#1}}
\newlength\fwidth
\newlength\swidth

\usepackage{pgfplots}
\usepackage{pgfplotstable}
\pgfplotsset{compat=1.5}
\usepackage{tikz}
\usepackage{pgfplots} 
\usetikzlibrary{decorations.text}
\usetikzlibrary{shapes.symbols}
\usetikzlibrary{arrows,calc,decorations.markings}
\usepgfplotslibrary{fillbetween}
\pgfplotsset{compat=1.5}
\usetikzlibrary{shapes,arrows}
\usetikzlibrary{positioning,shapes,shadows,arrows,intersections}
\pgfplotsset{compat=1.10}

\newcommand{\qscn}{\,{;}}
\newcommand{\zJ}{\bm J}
\newcommand{\zE}{\bm E}

\newcommand{\zp}{\bm p}

\newcommand{\zwel}{w_{\+{el},k}}
\newcommand{\zwth}{w_{\+{th},k}}
\newcommand{\zn}{\bm n}
\newcommand{\zr}{\bm r}
\newcommand{\qpt}{\,{.}}
\newcommand{\qcm}{\,{,}}
\newcommand{\idx}[1]{_\text{#1}}
\newcommand{\udV}{\,\text{d}\Omega}

\newcommand{\pd}[2]{\frac{\partial #1}{\partial#2}}
\newcommand{\dd}[2]{\frac{\text d #1}{\text d#2}}

\newcommand{\ttsigmad}{\ensuremath{\overline{\overline{\sigma}}_{\mathrm{d}}}}

\usepackage{aligned-overset}
\newcommand{\divergence}[1]{\,\text{div}\left(#1\right)}

\newcommand{\intV}[1]{ \int_\Omega #1 \udV}

\newcommand{\grad}[1]{\,\text{grad}\left(#1\right)}
\newcommand{\ttepsd}{\ensuremath{\overline{\overline{\varepsilon}}_{\mathrm{d}}}}
\allowdisplaybreaks

\newcommand{\zqdot}{\bm {\dot{q}}}
\newcommand{\qjoule}{\dot{q}_\+{Joule}}
\newcommand{\cv}{c\idx{V}}

%% file: introduction.tex
Cable joints are known to be the most vulnerable components of  \ac{hvdc} cable systems as they 
suffer from high internal electric field stresses \cite{CigreD156_2020aa,Chen_2015ac,Ghorbani_2014aa,Joergens_2020ab}. 
One solution to mitigate these stresses is the integration of a layer of nonlinear \ac{fgm} \cite{Hussain_2017aa, Secklehner_2017aa}. This material becomes highly conductive in areas of high stress, effectively reducing electric field stress and redistributing voltage drop to lower-stress areas.

Recent developments in material science allow for the customization of an \ac{fgm}'s nonlinear conductivity, enabling more precise designs to fit specific applications \cite{Bauer_2020aa, Secklehner_2017aa}. In addition to practical know-how, \ac{fe} simulations  play an increasingly vital role for the development of \acp{fgm}.  However, only few studies systematically investigated the various design parameters of \acp{fgm} \cite{Secklehner_2017aa,Hussain_2017aa,Joergens_2019ac,Spack_2021ab}. 

One way to study the influence of individual design parameters without extensive parameter sweeps is the use of sensitivities, i.e. gradients. Sensitivities provide insights on the impact of small changes in a design parameter on a \ac{qoi}, and allow an efficient optimization of \acp{fgm} \cite{Ion_2018aa}.

Different methods for sensitivity computation exist. Commonly used methods like, e.g., finite differences and the \acl{dsm}, 
have the drawback that their computational costs scale linearly with the number of parameters \cite{Li_2004aa, Nikolova_2004aa}. The \acl{am}, on the other hand, features computational costs that are nearly unaffected by the number of parameters \cite{Li_2004aa, Cao_2003aa, Nikolova_2004aa}. In \acl{hv} engineering, the \acl{am} has recently been formulated for linear \ac{eqs} problems in frequency domain \cite{Zhang_2021aa}, nonlinear \ac{eqs} problems in the time domain \cite{Ruppert_2023ac}, and stationary nonlinear coupled electrothermal problems \cite{Ruppert_2022ab}. 
However, for an investigation of the electrothermal behavior of a cable joint during transient overvoltages, a fully coupled transient \ac{eqst} analysis is required \cite{Hussain_2023aa, Koch_2018aa, Vu_2021aa}. This study formulates and numerically solves the \acl{am} for transient coupled \ac{eqst} problems with nonlinear material properties. The method is implemented within the Python-based FE framework \textit{Pyrit} \cite{Bundschuh_2023ab} and applied to the model of a 320\,kV \ac{hvdc} cable joint under switching impulse operation. The method is validated using results obtained via the \acl{dsm} as a reference. Moreover, the benefits of a multi-rate time-integration approach are demonstrated. The outcome of the paper is an \ac{fe}-based adjoint sensitivity analysis tool incorporating electrothermal multiphysics, nonlinear material properties and transient overvoltages, which are the three challenges to be tackled in cable joint design.

%% file: figs/ftsigma_RH.tex
%
%

\definecolor{mycolor1}{rgb}{0.24220,0.15040,0.66030}%
\definecolor{mycolor3}{rgb}{0.91840,0.73080,0.18900}%
\definecolor{mycolor2}{rgb}{0.10850,0.66690,0.87340}%
\begin{tikzpicture}
\begin{axis}[%
width=0.951\fwidth,
height=0.65\fwidth,
at={(0\fwidth,0\fwidth)},
scale only axis,
xmin=0,
xmax=2,
xlabel style={font=\color{white!15!black}},
xlabel={electric field strength $E$ in kV/mm},
ymode=log,
ymin=8e-11,
ymax=1e-3,
yminorticks=true,
ylabel style={font=\color{white!15!black}},
ylabel={conductivity $\sigma_{\+{FGM}}$ in S/m},
axis background/.style={fill=white},
xmajorgrids,
ymajorgrids,
yminorgrids,
legend style={legend cell align=left, align=left, draw=white!15!black, legend pos = north west}
]
\addplot [color=mycolor1, line width=2.0pt]
  table[row sep=crcr, x expr = \thisrowno{0} * 1e-6]{%
0	1.00053648068057e-10\\
30303.0303030303	1.00074324471879e-10\\
60606.0606060606	1.00102969730693e-10\\
90909.0909090909	1.00142655106334e-10\\
121212.121212121	1.0019763555004e-10\\
151515.151515152	1.00273805906031e-10\\
181818.181818182	1.00379332939655e-10\\
212121.212121212	1.00525530954362e-10\\
242424.242424242	1.00728074878608e-10\\
272727.272727273	1.01008680886359e-10\\
303030.303030303	1.01397434742481e-10\\
333333.333333333	1.01936017511452e-10\\
363636.363636364	1.02682174480582e-10\\
393939.393939394	1.03715906442733e-10\\
424242.424242424	1.05148047149983e-10\\
454545.454545455	1.07132146587977e-10\\
484848.484848485	1.09880934161241e-10\\
515151.515151515	1.13689126924503e-10\\
545454.545454545	1.18965028292447e-10\\
575757.575757576	1.26274305149882e-10\\
606060.606060606	1.3640063701682e-10\\
636363.636363636	1.50429739895365e-10\\
666666.666666667	1.69865773589879e-10\\
696969.696969697	1.96792613352084e-10\\
727272.727272727	2.34097276921649e-10\\
757575.757575758	2.85779462146335e-10\\
787878.787878788	3.5738038287735e-10\\
818181.818181818	4.56576881621245e-10\\
848484.848484849	5.94004517867331e-10\\
878787.878787879	7.8439788100742e-10\\
909090.909090909	1.0481703884532e-09\\
939393.939393939	1.41360294947594e-09\\
969696.96969697	1.91987615532311e-09\\
1000000	2.62127109509402e-09\\
1030303.03030303	3.59298919061209e-09\\
1060606.06060606	4.93921502446369e-09\\
1090909.09090909	6.80428661186021e-09\\
1121212.12121212	9.38817071747865e-09\\
1151515.15151515	1.29679023532892e-08\\
1181818.18181818	1.79272869795703e-08\\
1212121.21212121	2.47980496884915e-08\\
1242424.24242424	3.43168426818642e-08\\
1272727.27272727	4.7504222097393e-08\\
1303030.3030303	6.57740596957454e-08\\
1333333.33333333	9.1085116077733e-08\\
1363636.36363636	1.26151018736528e-07\\
1393939.39393939	1.74731142950922e-07\\
1424242.42424242	2.4203355369087e-07\\
1454545.45454545	3.35273154689478e-07\\
1484848.48484848	4.64444780124319e-07\\
1515151.51515152	6.43393910905357e-07\\
1545454.54545455	8.91299418109337e-07\\
1575757.57575758	1.23472654810059e-06\\
1606060.60606061	1.71046880458131e-06\\
1636363.63636364	2.3694806254977e-06\\
1666666.66666667	3.28231713421389e-06\\
1696969.6969697	4.54665390203097e-06\\
1727272.72727273	6.29767323485685e-06\\
1757575.75757576	8.72239277352653e-06\\
1787878.78787879	1.20794003409468e-05\\
1818181.81818182	1.67259726642688e-05\\
1848484.84848485	2.31552216205591e-05\\
1878787.87878788	3.20467480219687e-05\\
1909090.90909091	4.43352793368453e-05\\
1939393.93939394	6.13028454500743e-05\\
1969696.96969697	8.47009817882759e-05\\
2000000	0.000116909733797679\\
2030303.03030303	0.000161138842995848\\
2060606.06060606	0.000221671514268356\\
2090909.09090909	0.000304139349099001\\
2121212.12121212	0.000415793455560373\\
2151515.15151515	0.000565695301501934\\
2181818.18181818	0.000764687236809748\\
2212121.21212121	0.00102492256069253\\
2242424.24242424	0.00135867074244011\\
2272727.27272727	0.00177614354850854\\
2303030.3030303	0.0022823362271732\\
2333333.33333333	0.00287343736614203\\
2363636.36363636	0.00353410659302064\\
2393939.39393939	0.00423733693721845\\
2424242.42424242	0.00494801177869064\\
2454545.45454545	0.00562952083406273\\
2484848.48484848	0.00625097789376391\\
2515151.51515152	0.00679219578871147\\
2545454.54545455	0.00724497150071101\\
2575757.57575758	0.00761119661452739\\
2606060.60606061	0.00789941978432518\\
2636363.63636364	0.00812140779673517\\
2666666.66666667	0.00828955427041279\\
2696969.6969697	0.0084153161846427\\
2727272.72727273	0.00850848974583747\\
2757575.75757576	0.00857703568465424\\
2787878.78787879	0.00862720307676784\\
2818181.81818182	0.00866378057585047\\
2848484.84848485	0.00869037587028906\\
2878787.87878788	0.00870967428021159\\
2909090.90909091	0.00872365739177105\\
2939393.93939394	0.00873377845959663\\
2969696.96969697	0.00874109854260011\\
3000000	0.00874638987349994\\
};
\addlegendentry{$\theta =20^\circ$C}

\addplot [color=mycolor2, line width=2.0pt]
  table[row sep=crcr, x expr = \thisrowno{0} * 1e-6]{%
0	2.24700426506627e-10\\
30303.0303030303	2.2474686156269e-10\\
60606.0606060606	2.24811193068811e-10\\
90909.0909090909	2.24900318463083e-10\\
121212.121212121	2.25023793512608e-10\\
151515.151515152	2.25194856855212e-10\\
181818.181818182	2.25431849407919e-10\\
212121.212121212	2.25760180827048e-10\\
242424.242424242	2.26215053858098e-10\\
272727.272727273	2.26845238672341e-10\\
303030.303030303	2.27718301863571e-10\\
333333.333333333	2.28927850742935e-10\\
363636.363636364	2.30603569644163e-10\\
393939.393939394	2.32925124302828e-10\\
424242.424242424	2.36141425096955e-10\\
454545.454545455	2.40597314497864e-10\\
484848.484848485	2.46770540082487e-10\\
515151.515151515	2.5532297724642e-10\\
545454.545454545	2.67171593568512e-10\\
575757.575757576	2.83586763420224e-10\\
606060.606060606	3.06328473826435e-10\\
636363.636363636	3.37835025173471e-10\\
666666.666666667	3.81484458703208e-10\\
696969.696969697	4.41956740282862e-10\\
727272.727272727	5.25735532726948e-10\\
757575.757575758	6.41803355210365e-10\\
787878.787878788	8.0260466268076e-10\\
818181.818181818	1.02538010371771e-09\\
848484.848484849	1.33401501183509e-09\\
878787.878787879	1.76160031959427e-09\\
909090.909090909	2.35398046832681e-09\\
939393.939393939	3.17466870815356e-09\\
969696.96969697	4.31165678884162e-09\\
1000000	5.88684914973295e-09\\
1030303.03030303	8.06913310162441e-09\\
1060606.06060606	1.10924863214384e-08\\
1090909.09090909	1.52810630424824e-08\\
1121212.12121212	2.10839485122979e-08\\
1151515.15151515	2.91233078048123e-08\\
1181818.18181818	4.02610910066577e-08\\
1212121.21212121	5.56914460304974e-08\\
1242424.24242424	7.70687459764651e-08\\
1272727.27272727	1.06684955244102e-07\\
1303030.3030303	1.47715346237583e-07\\
1333333.33333333	2.0455890241154e-07\\
1363636.36363636	2.83309886862517e-07\\
1393939.39393939	3.92411102475306e-07\\
1424242.42424242	5.43558818627585e-07\\
1454545.45454545	7.52956262061572e-07\\
1484848.48484848	1.04304982574673e-06\\
1515151.51515152	1.44493368291642e-06\\
1545454.54545455	2.00167973143816e-06\\
1575757.57575758	2.77294818664222e-06\\
1606060.60606061	3.84136987843032e-06\\
1636363.63636364	5.32137825485749e-06\\
1666666.66666667	7.37142597225636e-06\\
1696969.6969697	1.0210872773669e-05\\
1727272.72727273	1.41433110056033e-05\\
1757575.75757576	1.9588744780566e-05\\
1787878.78787879	2.71279104856706e-05\\
1818181.81818182	3.75631799936266e-05\\
1848484.84848485	5.20019836803551e-05\\
1878787.87878788	7.19705686672083e-05\\
1909090.90909091	9.95681453764015e-05\\
1939393.93939394	0.000137673895801699\\
1969696.96969697	0.000190221417218188\\
2000000	0.000262555814349178\\
2030303.03030303	0.000361885522887908\\
2060606.06060606	0.000497829761955194\\
2090909.09090909	0.000683035978993071\\
2121212.12121212	0.000933788708428996\\
2151515.15151515	0.00127043818965819\\
2181818.18181818	0.00171733416595113\\
2212121.21212121	0.00230177050982914\\
2242424.24242424	0.00305130198851627\\
2272727.27272727	0.00398886218137087\\
2303030.3030303	0.00512566941415776\\
2333333.33333333	0.00645316401929703\\
2363636.36363636	0.00793689459710115\\
2393939.39393939	0.00951620890822119\\
2424242.42424242	0.0111122420671292\\
2454545.45454545	0.0126427747200323\\
2484848.48484848	0.0140384426348636\\
2515151.51515152	0.0152539094786614\\
2545454.54545455	0.0162707529177825\\
2575757.57575758	0.0170932210722271\\
2606060.60606061	0.0177405125047054\\
2636363.63636364	0.0182390530580089\\
2666666.66666667	0.018616676313937\\
2696969.6969697	0.0188991123501176\\
2727272.72727273	0.0191083614813973\\
2757575.75757576	0.0192623019122045\\
2787878.78787879	0.0193749678131718\\
2818181.81818182	0.0194571135400205\\
2848484.84848485	0.019516841237299\\
2878787.87878788	0.0195601815954391\\
2909090.90909091	0.0195915848595076\\
2939393.93939394	0.0196143147479327\\
2969696.96969697	0.0196307541862211\\
3000000	0.0196426374541772\\
};
\addlegendentry{$\theta =40^\circ$C}

\addplot [color=mycolor3, line width=2.0pt]
  table[row sep=crcr, x expr = \thisrowno{0} * 1e-6]{%
0	4.57917399681911e-10\\
30303.0303030303	4.58012029765428e-10\\
60606.0606060606	4.58143131056415e-10\\
90909.0909090909	4.5832475985626e-10\\
121212.121212121	4.58576389879768e-10\\
151515.151515152	4.58925000170562e-10\\
181818.181818182	4.59407967716136e-10\\
212121.212121212	4.60077074900394e-10\\
242424.242424242	4.61004061461133e-10\\
272727.272727273	4.62288316217314e-10\\
303030.303030303	4.64067533250863e-10\\
333333.333333333	4.66532475068008e-10\\
363636.363636364	4.69947425603606e-10\\
393939.393939394	4.74678525980419e-10\\
424242.424242424	4.81233031101477e-10\\
454545.454545455	4.90313696053726e-10\\
484848.484848485	5.02894123475732e-10\\
515151.515151515	5.203231504159e-10\\
545454.545454545	5.44469466737561e-10\\
575757.575757576	5.77921970636618e-10\\
606060.606060606	6.2426734280806e-10\\
636363.636363636	6.88474599955183e-10\\
666666.666666667	7.77427858345799e-10\\
696969.696969697	9.00664428762237e-10\\
727272.727272727	1.07139737921062e-09\\
757575.757575758	1.30793220152784e-09\\
787878.787878788	1.63562947263259e-09\\
818181.818181818	2.08962394099477e-09\\
848484.848484849	2.71859157035531e-09\\
878787.878787879	3.58996843116202e-09\\
909090.909090909	4.79718099211722e-09\\
939393.939393939	6.46966301884763e-09\\
969696.96969697	8.78673305503933e-09\\
1000000	1.19968203749087e-08\\
1030303.03030303	1.6444100730152e-08\\
1060606.06060606	2.2605397645609e-08\\
1090909.09090909	3.11413056111067e-08\\
1121212.12121212	4.29670162530555e-08\\
1151515.15151515	5.93504408845538e-08\\
1181818.18181818	8.2048149123481e-08\\
1212121.21212121	1.13493697129492e-07\\
1242424.24242424	1.57058534792091e-07\\
1272727.27272727	2.17413460446277e-07\\
1303030.3030303	3.01029367384101e-07\\
1333333.33333333	4.1687095182846e-07\\
1363636.36363636	5.77357723406165e-07\\
1393939.39393939	7.99695285164512e-07\\
1424242.42424242	1.10771948531559e-06\\
1454545.45454545	1.53445090851787e-06\\
1484848.48484848	2.12563309901206e-06\\
1515151.51515152	2.94463292785241e-06\\
1545454.54545455	4.07922669247412e-06\\
1575757.57575758	5.65099605203656e-06\\
1606060.60606061	7.82833452207691e-06\\
1636363.63636364	1.08444462303517e-05\\
1666666.66666667	1.50222421276258e-05\\
1696969.6969697	2.08087558252295e-05\\
1727272.72727273	2.88226787072316e-05\\
1757575.75757576	3.99199379031212e-05\\
1787878.78787879	5.52840171312963e-05\\
1818181.81818182	7.65500714612921e-05\\
1848484.84848485	0.000105974935230072\\
1878787.87878788	0.00014666895016661\\
1909090.90909091	0.000202910100931952\\
1939393.93939394	0.000280565877643019\\
1969696.96969697	0.000387652565197921\\
2000000	0.000535062962039357\\
2030303.03030303	0.00073748715211483\\
2060606.06060606	0.00101452815921586\\
2090909.09090909	0.00139196023902728\\
2121212.12121212	0.0019029696732842\\
2151515.15151515	0.00258902825112224\\
2181818.18181818	0.00349975835775307\\
2212121.21212121	0.00469078222463624\\
2242424.24242424	0.00621825376101104\\
2272727.27272727	0.00812890934912845\\
2303030.3030303	0.01044561083506\\
2333333.33333333	0.013150914456988\\
2363636.36363636	0.0161746116460833\\
2393939.39393939	0.0193930990956707\\
2424242.42424242	0.0226456579149648\\
2454545.45454545	0.0257647331363239\\
2484848.48484848	0.0286089672675848\\
2515151.51515152	0.0310859693149977\\
2545454.54545455	0.0331581963719954\\
2575757.57575758	0.0348343056898979\\
2606060.60606061	0.036153422053872\\
2636363.63636364	0.0371693987360435\\
2666666.66666667	0.0379389578423718\\
2696969.6969697	0.0385145347439157\\
2727272.72727273	0.0389409639215144\\
2757575.75757576	0.0392546793998384\\
2787878.78787879	0.0394842814402344\\
2818181.81818182	0.0396516863633951\\
2848484.84848485	0.0397734055441371\\
2878787.87878788	0.0398617289372403\\
2909090.90909091	0.0399257257050585\\
2939393.93939394	0.0399720469851935\\
2969696.96969697	0.040005548945783\\
3000000	0.0400297658787313\\
};
\addlegendentry{$\theta =60^\circ$C}

\end{axis}
\end{tikzpicture}%

%% file: electrothermal_modeling.tex
The capacitive-resistive-thermal behavior of a cable joint that is subjected to transient overvoltages, e.g. a lightning strike or switching operations, can be described by the combination of the transient \ac{eqs} equation and the transient heat conduction equation \cite{Saltzer_2011aa}. The transient \ac{eqs} problem reads
\begingroup
   \begin{subequations}\label{eq:eqs}
     \begin{alignat}{2}
     \label{eq:eqs:1} \divergence{\bm J + \partial_t{\bm D}}&=0,&& \quad\quad t \in [t_\+{s},t_\+{f}],\,\, \zr \in\Omega\qscn\\
      \label{eq:eqs:5}\phi &=\phi_\text{D},\,  && \quad\quad  t \in [t_\+{s},t_\+{f}],\,\, \zr \in\Gamma_\text{D,el} \qscn\\
      \label{eq:eqs:6}\left(\bm J + \partial_t{\bm{D}}\right) \cdot \bm n\idx{el} &=0,  && \quad\quad  t \in [t_\+{s},t_\+{f}],\,\,\zr \in\Gamma_\text{N,el} \qscn\\
      \label{eq:eqs:7}\phi &=\phi_0, && \quad\quad  t  = t_\+{s}, \,\,\zr \in\Omega \qcm
     \end{alignat}
   \end{subequations}
\endgroup
  where $\bm J$ is the resistive current density, $\bm D$ is the electric displacement field, $\zE$ is the electric field and $\phi$ is the electric scalar potential. $\sigma$ represents the electric conductivity and $\varepsilon$ represents the electric permittivity. $\zr$ and $\Omega$ denote the spatial coordinate and the computational domain in space, respectively. $t$ is the time and $t_\+{s}$ and $t_\+{f}$ denote the initial and final simulation time, respectively. $\phi\idx{D}$ are the fixed voltages at the Dirichlet boundaries, $\Gamma\idx{\+{D,\,el}}\neq\emptyset$. The Neumann boundaries are denoted as $\Gamma_\text{N,el} =\partial\Omega \setminus \Gamma_\text{D,el}$. $\phi_0$ denotes the initial condition of the electric potential, i.e. the steady state potential distribution before the transient event.

The transient heat conduction equation reads
\begingroup
\begin{subequations}\label{eq:thc}
\begin{alignat}{3}
  \label{eq:thc:1}\partial_t\left(c_{\+{V}}  \theta\right) +\divergence{\zqdot} &= \dot q_\+{Joule},
  && \quad\quad t \in [t_\+{s},t_\+{f}] ,\, \bm r \in \Omega  \qscn\\
 \label{eq:thc:3} \theta &=\theta\idx{D},
  &&\quad\quad t \in [t_\+{s},t_\+{f}],\, \zr \in \Gamma\idx{\+{D,th}}  \qscn\\
  \label{eq:thc:4}\zqdot\cdot\zn_\+{th} &=0,
  && \quad\quad t \in [t_\+{s},t_\+{f}]  ,\,\zr \in \Gamma\idx{\+{N,th}}  \qscn\\
 \label{eq:thc:5}  \theta &=\theta_0,
  && \quad\quad t = t_\+{s},\,\zr \in \Omega \qcm
\end{alignat}
\end{subequations}
\endgroup
where $\zqdot=-\lambda\grad{\theta}$ is the heat flux density, $c_\+{V}$ is the volumetric heat capacity and $\lambda$ is the thermal conductivity. $\theta_0$ denotes the initial condition of the temperature. $T\idx{D}$ are the fixed temperatures at the Dirichlet boundaries, $\Gamma\idx{\+{D,\,th}}\neq\emptyset$. At the Neumann boundaries, $\Gamma_\text{N,th} =\partial\Omega \setminus \Gamma_\text{D,th}$, the normal heat flux density is set to zero. The two equations are coupled along the Joule losses, $\dot q_\+{Joule} = \bm J \cdot \bm E$, and possible temperature dependencies of electric materials, e.g., the \ac{fgm} conductivity defined in \eqref{Conductivity}.

%% file: eqst_numerical_approach.tex
\begin{figure}[tbh]	
	\centering
	\includegraphics[width=0.8\columnwidth]{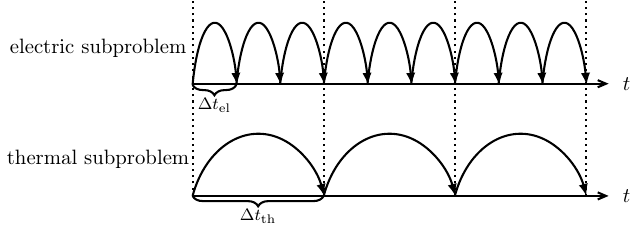}
	\caption{Illustration of the weak multi-rate coupling scheme: With thermal dynamics spanning minutes to hours and electric phenomena occuring on the microsecond to millisecond scale, distinct time step sizes are adopted for the electric and thermal subproblems. The thermal time step, $\Delta t\idx{th}$, is set as a multiple of the smaller electric time step, $\Delta t\idx{el}$. As indicated by the dotted lines, the field distributions are exchanged between both subproblems after each thermal time step.}
	\label{fig:mr_scheme}
\end{figure}

The electrothermal behavior of the cable joint is formulated as a \ac{2d} axisymmetric \ac{fe} problem. The electric scalar potential and the temperature are discretized by
\begin{subequations}
\begin{alignat}{1}
 \label{eq:phidisc} \phi(\zr,t) &\approx \sum_r u_r(t) N_r(\zr)\qcm\\
   \label{eq:tempdisc}\theta(\zr,t) &\approx \sum_r v_r(t) N_r(\zr)
\end{alignat}
\end{subequations}
where $N_r$ are linear nodal FE shape functions and $u_r$ and $v_r$ are the \acl{dof}, which are assembled in the vectors $\mathbf{u}$ and $\mathbf{v}$, respectively. The semi-discrete versions according to the Ritz-Galerkin procedure of \eqref{eq:eqs} and \eqref{eq:thc} read 
\begin{subequations}
\begin{alignat}{1}
\label{eq:eqsfemsemi} \mathbf{K}_\sigma \mathbf{u} + \partial_t\left(\mathbf{K}_\varepsilon \mathbf{u}\right) &= 0\qcm\\
\label{eq:thcfemsemi} \mathbf{K}_\lambda\mathbf{v} + \partial_t\left(\mathbf{M}_{c\idx{V}} \mathbf{v}\right)&= \bm{s}_{\qjoule},
\end{alignat}
\end{subequations}
with 
\begin{subequations}
\begin{alignat}{2}
  \label{eq:bla1} [\bm K_{(\cdot)}]_{rs} &=\int_\Omega (\cdot)\grad {N_r}\cdot \grad{N_s}\udV &&\qscn\\
   \label{eq:bla2}   [\bm M_{(\cdot)}]_{rs}  &=\int_\Omega {(\cdot)}{N_r}{N_s}\udV\qscn&&\qscn \\
   \label{eq:bla3}[\bm s_{(\cdot)}]_r &= \int_\Omega {(\cdot)}{N_r}\udV\qscn&&\qpt
\end{alignat}
\end{subequations}
The model is implemented for axisymmetric problems in the freely available \ac{fe} framework \textit{Pyrit} \cite{Bundschuh_2023ab}. The integrals \eqref{eq:bla1}--\eqref{eq:bla3} are carried out on a triangular mesh of the cut shown in Fig.~\ref{Muffe Konfiguration}, i.e., $\text{d}\Omega = 2\pi\varrho\text{d}\varrho\text{d}z$. The matrices are assembled for $r,s = 1,...,N\idx{node}$, where $N\idx{node}$ denotes the number of nodes of the mesh. A weak multi-rate coupling scheme (see Fig.~\ref{fig:mr_scheme}) \cite{Schierz_2012ac}
with implicit Euler time-stepping is performed. The nonlinearity arising in the electric subproblem \eqref{eq:eqsfemsemi} due to the \ac{fgm} is solved using the Newton method.

%% file: sensitivities_general.tex
As shown in Sec.~\ref{sec:specimen}, the nonlinear conductivity curve of an \ac{fgm} is represented as an analytical function shaped by various design parameters, denoted as $\bm p =[p_1,...,p_{N_{\+{P}}}]$.  When designing an \ac{fgm}, the quality of the nonlinear conductivity curve is evaluated based on a number of  \acp{qoi}, $G_k$, $k=1,...,N_{\+{QoI}}$, such as the Joule losses or the electric field at critical positions \cite{Secklehner_2017aa, Hussain_2017aa}.   The design process, i.e. the optimization of $\bm p$, can be accelerated by using sensitivity information \cite{Ion_2018aa}. Sensitivities quantify the influence of small changes in a design parameter, $p_j$, on a given QoI, $G_k$, expressed as $\dd {G_k} {p_j}(\bm{p}_0)$, where $\bm{p}_0$ represents the current parameter configuration. In this section, it is discussed how sensitivities of \ac{hvdc} cable joint materials can be computed most effectively. 

\subsection{Direct Sensitivity Method}
One of the most frequently used methods for sensitivity computation is the \ac{dsm}. The \ac{dsm} appeals with its simple derivation and uncomplicated implementation. It is based on the derivatives of \eqref{eq:eqs} and \eqref{eq:thc} to $p_j$, 
from which the derivatives of the electric potential and the temperature to $p_j$ can be computed, i.e. $\dd{\phi}{p_j}(\bm p_0)$,  and $\dd{\theta}{p_j}(\bm p_0)$. The sensitivity can then be calculated directly using the chain rule:
\begin{equation}
\label{eq:chain}
\dd {G_k} {p_j}(\bm p_0) = \pd{G_k}{p_j}(\bm p_0) + \pd{G_k}{\phi}\dd{\phi}{p_j}(\bm p_0)+ \pd{G_k}{\theta}\dd{\theta}{p_j}(\bm p_0)\qpt
\end{equation}
Since this process has to be repeated for each parameter, $p_1,...,p_{N_{\+{P}}}$, the \ac{dsm} requires the solution of $N_\+{P}$ individually coupled systems of linear \acp{pde} in addition to the nominal solution. Consequently, for large numbers of parameters, the \ac{dsm} leads to prohibitively long simulation times.

\subsection{Adjoint Variable Method}
An alternative approach for sensitivity computation is the \ac{am}. The \ac{am} is very efficient when the number of parameters, $N_{\+{P}}$, is greater than the number of quantities of interest (\acp{qoi}), $N_{\+{QoI}}$. It has originally been applied for the analysis of electric networks and only recently gained interest in the \ac{hv} engineering community \cite{Zhang_2021aa, Ruppert_2022ab, Ruppert_2023ac}. In this paper, the method is adapted for the \textit{transient coupled electrothermal} analysis of HVDC cable joints.

The \ac{am} avoids the computation of  $\dd{\phi}{p_j}(\bm p_0)$ and $\dd{\theta}{p_j}(\bm p_0)$ for each parameter by a clever representation of the \acp{qoi}: Each \ac{qoi}, $G_k$, is formulated as an integral over the computational domain in space and time, $\Omega \times [t\idx{s}, t_\+{f}]$, by means of a functional, $g_k$. Furthermore, the \ac{eqs} equation \eqref{eq:eqs} and the transient heat conduction equation \eqref{eq:thc} are added, multiplied by test functions $\zwel$ and $\zwth$, respectively as additional terms:
\begingroup 
\begin{align}
 \nonumber G_k&(\phi, \theta, \bm p) = \int_\Omega \int_{t_\+{s}}^{t_\+{f}} g_k(\phi, \theta,\zr, t, \bm p)\+{d}t\,\+d\Omega \\ 
\nonumber & -  \int_\Omega \int_{t_\+{s}}^{t_\+{f}} w_{\+{el},k}(\bm r, t) \underbrace{\left(\divergence{\partial_t \bm D + \bm J}\right)}_{ _{\overset{\+{Eq.}\eqref{eq:eqs:1}}{=} 0}} \+{d}t\,\+d\Omega\\
& -  \int_\Omega \int_{t_\+{s}}^{t_\+{f}} w_{\+{th},k}(\bm r, t) \underbrace{\left(\partial_t\left(c_{\+{V}}  \theta\right) +\divergence{\zqdot} - \dot{q}_\+{Joule}\right)}_{ _{\overset{\+{Eq.}\eqref{eq:thc:1}}{=} 0}} \+{d}t\,\+d\Omega\qpt\label{eq:qoi:fctl}
\end{align}
\endgroup
As indicated by the curved brackets, the additional terms are zero by construction. Consequently, the test functions, $\zwel$ and $\zwth$, can be chosen freely without changing the value of the \acp{qoi}. Taking the derivative of \eqref{eq:qoi:fctl} to $p_j$ yields:
\begingroup 
\begin{align}
\label{eq:sens_ext}
\nonumber\dd {G_k} {p_j}&(\bm p_0) = \int_\Omega \int_{t_\+{s}}^{t_\+{f}} \pd{g_k}{p_j}(\bm p_0) \+{d}t\,\+d\Omega 
+\int_\Omega \int_{t_\+{s}}^{t_\+{f}} \pd{g_k}{\phi}\dd{\phi}{p_j}(\bm p_0)+ \pd{g_k}{\theta}\dd{\theta}{p_j}(\bm p_0)\+{d}t\,\+d\Omega \\
-&  \nonumber\int_\Omega \int_{t_\+{s}}^{t_\+{f}} w_{\+{el},k}(\bm r, t)\cdot f_\+{el}\left(\dd{\phi}{p_j}, \dd{\theta}{p_j},\bm r, t, \bm p_0\right)\+{d}t\,\+d\Omega \\
-&  \int_\Omega \int_{t_\+{s}}^{t_\+{f}} w_{\+{th},k}(\bm r, t)\cdot f_\+{th}\left(\dd{\phi}{p_j}, \dd{\theta}{p_j},\bm r, t, \bm p_0\right)\udV\+{d}t\qcm
\end{align}
\endgroup
where $f_\+{el}$ and $f_\+{th}$ are the derivatives of \eqref{eq:eqs:1} and \eqref{eq:thc:1} to $p_j$, respectively, i.e.
\begingroup 
\begin{subequations}
\begin{alignat}{1}
f_\+{el} &= \dd{}{p_j}\left(\divergence{\partial_t\bm D + \bm J}\right)(\bm p_0)\qcm\\
f_\+{th} &= \dd{}{p_j}\left(\partial_t\left(c_{\+{V}}  \theta\right) + \divergence{\zqdot} - \dot{q}_\+{Joule}\right)(\bm p_0)\qpt
\end{alignat}
\end{subequations}
\endgroup
Equation~\eqref{eq:sens_ext} shows that the test functions are multiplied by factors containing $\dd{\phi}{p_j}(\bm p_0)$ and $\dd{\theta}{p_j}(\bm p_0)$. The idea of the \ac{am} is to choose the test functions in such a way that all occurrences of these unwanted terms vanish. This is achieved by choosing the test functions as the solutions of the so-called adjoint problem \cite{Li_2004aa, Cao_2003aa}. The adjoint problem for nonlinear coupled \ac{eqst} problems with parameter-dependent materials, i.e., $\sigma(\zE, \theta, \zp)$, $\varepsilon(\zE, \theta, \zp)$, $\lambda(\theta, \zp)$, $c_{\+{V}}(\theta, \zp)$, reads:

\begingroup 
\begin{subequations}
   \label{eq:eqs:adj}
   \begin{alignat}{3}
      \+{div}\left({\ttepsd{\pd{}{t}\grad{\zwel}}}\right) -\+d\+{iv}\left({\ttsigmad\grad \zwel}\right)&&&&\nonumber\\
     \phantom{=}+\divergence{\zwth\left({\ttsigmad \zE + \zJ}\right)} &=\pd{g_k}{\phi}&&\qcm \quad t \in [t_\+{s},t_\+{f}]&&,\,\, \zr \in\Omega \qscn\\
   \zwel &=0&&\qcm\quad t \in [t_\+{s},t_\+{f}]&&,\,\, \zr \in\Gamma\idx{D,el} \qscn\\
\hspace{-1mm}\left(-{\ttsigmad\grad \zwel}+{\ttepsd{\pd{}{t}\grad{\zwel}}}+{\zwth\left({\ttsigmad \zE + \zJ}\right)}\right)\cdot \bm n\idx{el} &=0&&\qcm \quad t \in [t_\+{s},t_\+{f}]&&,\,\, \zr \in\Gamma\idx{N,el} \qscn\\
     \zwel &=0&&\qcm\quad t= t_\+{f}&&,\,\, \zr \in\Omega \qcm \label{eq:adjel:term}
     \end{alignat}
   \end{subequations}
  \endgroup
   and
   \begingroup
   \begin{subequations}
   \label{eq:thc:adj}
   \begin{alignat}{3}
      \pd{}{t}\grad{\zwel}\cdot \pd{\varepsilon}{\theta}\zE-\grad \zwel \cdot \pd{\sigma}{\theta}\zE&\nonumber\\
      - \pd{\zwth}{t}\left(\cv+\pd{\cv}{\theta}\theta\right)-\divergence{\lambda\grad{\zwth}}&\nonumber\\
     + \grad \zwth \cdot \pd{\lambda}{\theta}\grad{\theta}-\zwth \pd{\sigma}{\theta}E^2&=\pd{g_k}{\theta}&&\qcm\quad t \in [t_\+{s},t_\+{f}]&&,\,\, \zr \in\Omega \qscn\\
   \zwth &=0&&\qcm\quad t \in [t_\+{s},t_\+{f}]&&,\,\, \zr \in\Gamma\idx{D,th} \qscn\\
-\lambda\grad \zwth\cdot \bm n\idx{th} &=0&&\qcm\quad t \in [t_\+{s},t_\+{f}]&&,\,\, \zr \in\Gamma\idx{N,th} \qscn\\
     \zwth &=0 &&\qcm\quad t= t_\+{f}&&,\,\, \zr \in\Omega \qcm \label{eq:adjth:term}
     \end{alignat}
   \end{subequations}
  \endgroup
  where all quantities are evaluated at the currently implemented parameter configuration, $\bm p_0$. Moreover, $\ttsigmad$ and $\ttepsd$ denote the differential electric conductivity and differential permittivity, respectively, i.e. \cite{Ruppert_2023ac, De-Gersem_2008aa}
 \begin{align*}
  \ttsigmad(\zE,\theta) &=\sigma(\zE, \theta) \left[\begin{array}{ccc}
    1 & 0    \\  0& 1
  \end{array}\right] +2\dd{\sigma}{E^2}(\zE, \theta)\zE \zE^T
  \qcm \\
   \ttepsd(\zE,\theta) &=\varepsilon(\zE, \theta) \left[\begin{array}{ccc}
    1 & 0    \\  0& 1
  \end{array}\right] +2\dd{\varepsilon}{E^2}(\zE, \theta)
  \zE \zE^T \qcm \\
\end{align*}
and are evaluated for the operating points defined by the nominal solution. Once the electric potential, the temperature and the test functions are available, the sensitivity with respect to any parameter can be computed directly by
 \begin{align}
  &\dd {G_k} {p_j}(\bm p_0)=\int_{t_\+{s}}^{t_\+{f}}\int_\Omega\pd{g}{p_j} \udV\+dt\nonumber\\
 &\phantom{=}  +\int_{t_\+{s}}^{t_\+{f}}\int_\Omega \grad \zwel \cdot \pd{\sigma}{p_j}\zE  \udV\+dt \nonumber\\
  &\phantom{=}  -\int_{t_\+{s}}^{t_\+{f}}\int_\Omega   \pd{}{t}\grad{\zwel} \cdot \pd{\varepsilon}{p_j}\zE \udV\+dt \nonumber\\
   &\phantom{=} + \int_{t_\+{s}}^{t_\+{f}}\int_\Omega \pd{\zwth}{t}\pd{\cv}{p_j}\theta\udV\+dt\nonumber\\
   &\phantom{=} + \int_{t_\+{s}}^{t_\+{f}}\int_\Omega\zwth\pd{\sigma}{p_j}E^2\udV\+dt\label{eq:adjoint_formulation2}\nonumber\\
   &\phantom{=} - \int_{t_\+{s}}^{t_\+{f}}\int_\Omega- \grad \zwth \cdot \pd{\lambda}{p_j}\grad{\theta}\udV\+dt\nonumber\\
   &\phantom{=} - \intV{\grad \zwel\cdot {\dd{\bm D}{p_j}}}\bigg\rvert_{t = t_\+{s}}\nonumber\\
  &\phantom{=}{+ \intV{\zwth\left(\pd{\cv}{p_j}+\pd{\cv}{\theta}\dd{\theta}{p_j}\right)\theta }\bigg\rvert_{t = t_\+{s}}}\nonumber\\
    &\phantom{=}{+ \intV{\zwth \cv\dd{\theta}{p_j}}\bigg\rvert_{t = t_\+{s}}}\qcm
\end{align} 
where the integrals evaluated at $t=0$ are computed by differentiating the initial conditions \eqref{eq:eqs:7} and \eqref{eq:thc:5}. Since the adjoint problem \eqref{eq:eqs:adj} and \eqref{eq:thc:adj} is independent of the choice of $p_j$, the same test functions, $\zwel$ and $\zwth$, can be used to compute the derivative of $G_k$ with respect to any parameter. Hence, the \ac{am} requires the solution of only one additional coupled linear system of \acp{pde} per \ac{qoi} independently of the number of parameters $N_\+{P}$. 

%% file: avm_eqst_numerical_approach.tex
The electric scalar potential and the temperature are discretized in space according to \eqref{eq:phidisc} and \eqref{eq:tempdisc}, respectively. Furthermore, the following spatial discretizations are applied:
\begin{subequations}
\begin{alignat}{1}
 \label{eq:jstat:discretephiprime} \dd{\phi}{p_j}(\zr) &\approx\sum_r u'_r(t) N_r(\zr)\qscn\\
  \label{eq:jstat:discretethetaprime} \dd{\theta}{p_j}(\zr) &\approx\sum_r v'_r(t) N_r(\zr)\qscn\\
  \label{eq:jstat:discreteadj} \zwel(\zr) &\approx\sum_r w_{\+{el},j}(t) N_r(\zr)\qscn\\
  \label{eq:jstat:discreteadj} \zwth(\zr) &\approx\sum_r w_{\+{th},j}(t) N_r(\zr) \qpt
\end{alignat}
\end{subequations}
The semi-discrete counterpart of the adjoint formulation, \eqref{eq:eqs:adj} and \eqref{eq:thc:adj}, reads 
\begin{subequations}
\begin{alignat}{1}
 -\bm K_{\ttepsd} \pd{}{t} \bm \zwel 
 + \bm K_{\ttsigmad}  \bm \zwel - \bm A^T_{\bm J + \ttsigmad \bm E} \bm \zwth 
 &=  \bm x_{\+{el}} \qscn \label{eq:jstat:adjel:disc}\\
   \bm A_{\pd{\varepsilon}{\theta}\zE} \pd{}{t}\bm \zwel 
   - \bm A_{\pd{\sigma}{\theta}\zE} \bm \zwel  - \bm M_{\cv + \pd{\cv}{\theta}\theta} \pd{}{t}\bm\zwth\nonumber&\\
   \phantom{=}+  \left(\bm K_{\lambda} -  \bm A_{-\pd{\lambda}{\theta}\grad{\theta}} - \bm M_{\pd{\sigma}{\theta}E^2}\right) \bm \zwth&=  \bm x_{\+{th}} \label{eq:jstat:adjth:disc}\qcm
\end{alignat}
\end{subequations}
with
\begin{subequations}
\begin{alignat}{1}
[\bm A_{\bm{(\cdot)}}]_{rs}   &=\int_\Omega\bm{(\cdot)} \cdot \grad{N_s} N_r\udV \qscn\\
 [\bm x\idx{el}]_{j} &=  \intV{\pd{g_k}{u_j}} \qscn\\
  [\bm x\idx{el}]_{j} &=  \intV{\pd{g_k}{v_j}} \qpt
\end{alignat}
\end{subequations}
The adjoint problem is also implemented in \textit{Pyrit}. A multi-rate coupling scheme with implicit Euler time-stepping is performed. Since the adjoint problem provides terminal conditions instead of initial conditions (see \eqref{eq:adjel:term}, \eqref{eq:adjth:term}), the time-stepping is performed backwards in time. The coupling between \eqref{eq:jstat:adjel:disc} and \eqref{eq:jstat:adjth:disc} is resolved using the successive substitution method. 
The semi-discrete counterpart of \eqref{eq:adjoint_formulation2} is given by
 \begin{align}
  &\dd {G_k} {p_j}(\bm p_0)=\int_0^{t_\+{f}}\int_\Omega\pd{g}{p_j} \udV\+dt\nonumber\\
   &\phantom{=} - \int_0^{t_\+{f}} \bm u^T K_{\pd{\sigma}{p_j}} \bm \zwel \+dt\nonumber\\
    &\phantom{=} + \int_0^{t_\+{f}} \bm u^T K_{\pd{\varepsilon}{p_j}} \pd{}{t}\bm \zwel\+dt\nonumber\\
   &\phantom{=} + \int_0^{t_\+{f}}\bm v^T\bm M_{\pd{\cv}{p_j}}\pd{}{t}\bm \zwth\+dt\nonumber\\
   &\phantom{=} + \int_0^{t_\+{f}}\pd{}{t}\bm \zwth^T \bm s_{\pd{\sigma}{p_j}E^2}\+dt\nonumber\\
   &\phantom{=} + \int_0^{t_\+{f}}-\bm v^TK_{\pd{\lambda}{p_j}} \bm \zwth  \+dt\nonumber\\
   &\phantom{=} + \bm u^T K_{\pd{\varepsilon}{p_j}} \bm \zwel\bigg\rvert_{t = t_\+{s}}  + (\bm u')^T K_{\ttepsd} \bm\zwel \bigg\rvert_{t = t_\+{s}} \nonumber\\
   &\phantom{=}+ \bm u^T K_{\pd{\varepsilon}{\theta}\dd{T}{p}} \bm \zwel\bigg\rvert_{t = t_\+{s}} \nonumber\\
   &\phantom{=} + \bm v^T \bm M_{\pd{\cv}{p_j}+\pd{\cv}{T}\dd{T}{p_j}}\bm\zwth\bigg\rvert_{t = t_\+{s}} \nonumber\\
   &\phantom{=}+ (\bm v')^T \bm M_{\cv}\bm\zwth\bigg\rvert_{t = t_\+{s}}
\end{align} 

%% file: results.tex
In this section, the \ac{am} presented in the previous section is validated.
The method is applied to the 320\,kV cable joint model and the results are compared to results obtained via the \ac{dsm}.

The cable joint is investigated during a switching overvoltage event, which is defined according \cite{Kuchler_2018aa, CigreB132_2012aa} as
\begin{align}
\begin{split}
&U_\+{switch}(t) = U\idx{DC} + \hat U \, \frac{\tau_2}{\tau_2-\tau_1}\left( \+{exp}\left( -\frac{t}{\tau_2}\right)-\+{exp}\left( -\frac{t}{\tau_1}\right)\right)\qcm
\end{split}
\end{align}
with $U\idx{DC} = 320\,$kV, $\hat U= 1.15\,U\idx{DC}=368\,\+k\+V$ and the constants $\tau_1 = \frac{250}{2.41}\,\upmu\+s$ and $\tau_2 = \frac{2500}{0.87}\,\upmu\+s$. The impulse over the simulated time span $[t\idx{s},t\idx{f}]=[0, 30\,\+{ms}]$ is shown in Fig.~\ref{fig:switching_impulse}. 
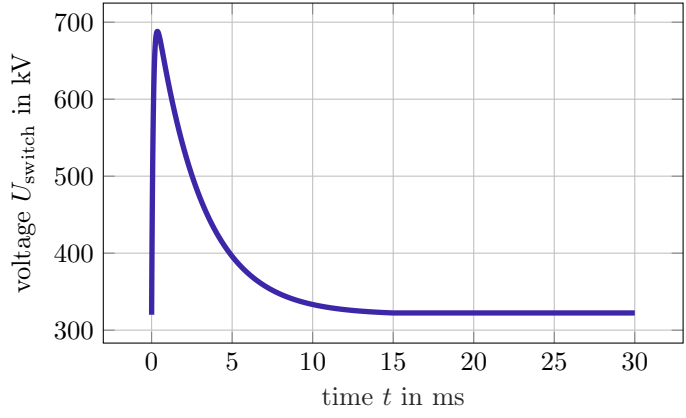
\begin{figure}[tb]	
	\centering
	\setlength{\fwidth}{0.45\columnwidth}
	\input{figs/switching_impulse.tex}\label{fig:switching_impulse}
	\caption{The switching impulse over the simulated time span $[0, 30\,\+{ms}]$. }
	\label{fig:switching_impulse}
\end{figure}
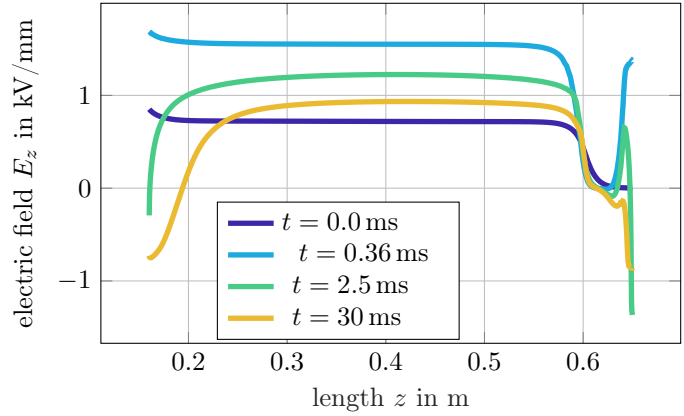
\begin{figure}[tb]	
	\centering
	\setlength{\fwidth}{0.45\columnwidth}
	\input{figs/switching_Ez_z_t.tex}\label{fig:switching_Ez_z_t}
	\caption{The tangential electric field strength, $E_z$, along the interface of the \ac{xlpe} and the \ac{fgm} (red line in Fig.~\ref{Muffe Konfiguration}) for different time instances.}
	\label{fig:switching_Ez_z_t}
\end{figure}
Figure~\ref{fig:switching_Ez_z_t} shows the tangential electric field strength along the interface between the \ac{xlpe} and the \ac{fgm} for several time instances. The Joule heat is 35.6\,J and no significant temperature rise is observed. 

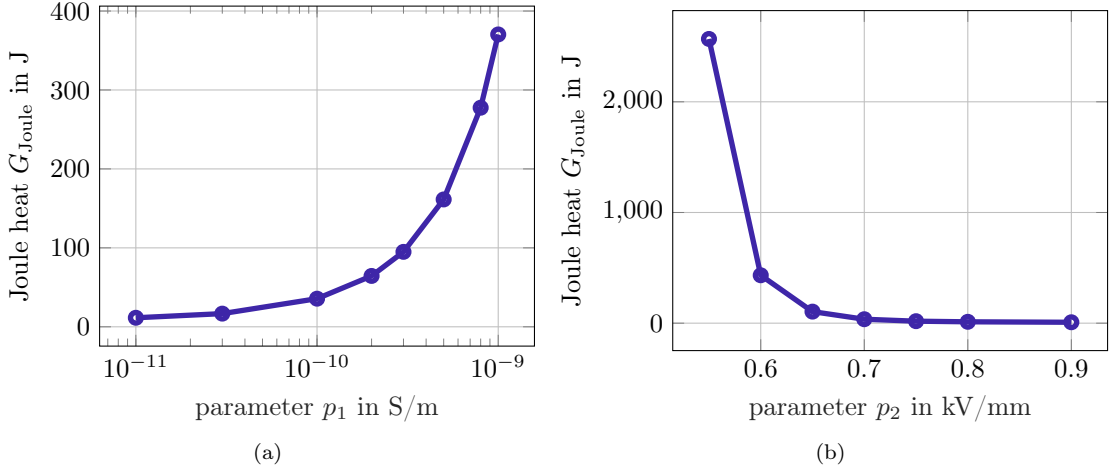
\begin{figure}[tb]
	\setlength{\fwidth}{0.45\columnwidth}	
	\centering
	\subfloat[]{
		\label{fig:sweep_wloss_sigma0}\input{figs/sweep_wloss_sigma0.tex}}
		\subfloat[]{
		\label{fig:sweep_wloss_E1}\input{figs/sweep_wloss_E1.tex}}	
		\caption{Joule heat as an exemplary \ac{qoi} for (a) different values of $p_1$ and (b) different values of $p_2$.}
	\label{fig:sweep}
\end{figure}

\begin{figure}[tb]
	\setlength{\fwidth}{0.45\columnwidth}	
	\centering
	\subfloat[]{
		\label{fig:sweep_wloss_normalized_sigma0}\input{figs/sweep_wloss_sigma0_normalized.tex}}
		\subfloat[]{
		\label{fig:sweep_wloss_normalized_E1}\input{figs/sweep_wloss_E1_normalized.tex}}	
		\caption{The \ac{qoi}, i.e. the Joule heat, for different values of (ab) $p_1$ and (b) $p_2$. The red lines indicate the tangent slope computed by the \ac{am}.}
	\label{fig:simple:convergence}
\end{figure}
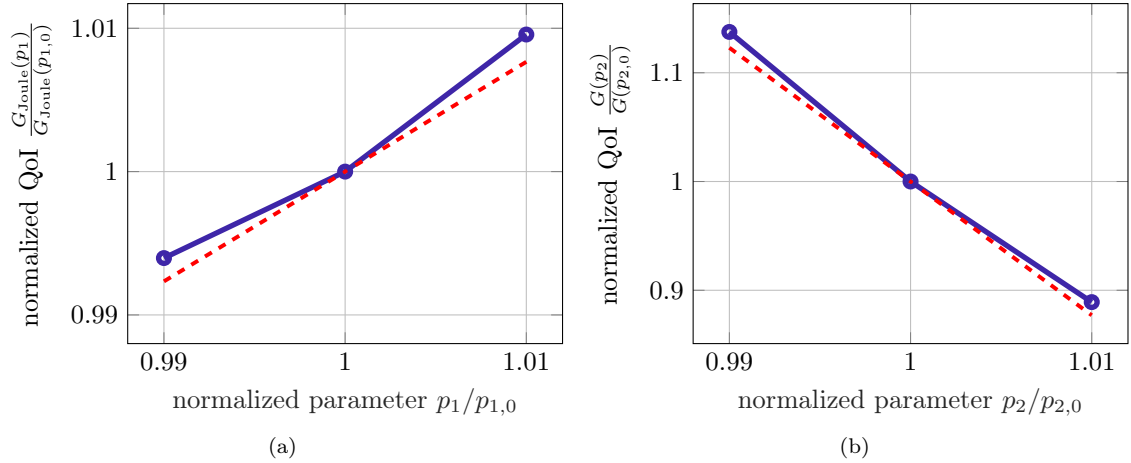

The validation is performed by investigating the sensitivity of the Joule heat with respect to the material parameters of the \ac{fgm}, i.e. $p_1$ to $p_5$ of \eqref{Conductivity}. This is motivated by the findings of \cite{Spack_2021ab} and \cite{Hussain_2017aa}, where it has been demonstrated that inappropriate choices for $p_1$ to $p_5$ can result in a substantial elevation of the Joule heat and consequently a significant temperature increase (see Fig.~\ref{fig:sweep}).
To apply the \ac{am}, the Joule heat is written in terms of a functional $g\idx{Joule}$,
\begin{align}
&G\idx{Joule}(\phi, T, \bm p) = \int_\Omega \int_{t\idx{s}}^{t\idx{f}} g\idx{Joule}(\phi, T,\zr, t, \bm p)\udV\,\+{d}t  = \int_\Omega \int_{t\idx{s}}^{t\idx{f}} \bm J \cdot \bm E\udV\,\+{d}t \qpt
\end{align}
Note that this integral notation does not restrict the choice of \acp{qoi}. \acp{qoi} that may not inherently be expressed as an integral can be effectively represented using Dirac delta functions, $\delta$, within the integral expression. For example, the temperature $\theta(\zr\idx{QoI},t\idx{QoI})$ at a particular position, $\zr\idx{QoI}$, and time instance, $t\idx{QoI}$, reads in integral notation:
\begin{equation}
G_{\theta(\zr\idx{QoI},t\idx{QoI})}(\phi, T, \bm p) = \int_\Omega \int_{t\idx{s}}^{t\idx{f}} \theta(\zr,t) \delta(\zr - \zr\idx{QoI})\delta(t - t\idx{QoI}) \udV\,\+{d}t \qpt
\end{equation}
For more information on \acp{qoi} that are evaluated at specific points in space or time and the numeric implications, see \cite{Ruppert_2023ac}. 
The \ac{qoi}-dependent parts of the \ac{fe} formulation are defined by
\begin{subequations}
\begin{alignat}{1}
\int_\Omega\pd{g\idx{Joule}}{p_j} \udV &\approx \bm u^T \bm K_{\pd{\sigma}{p}}\bm u\qcm\\
\bm{x}_{\+{el}} &= \bm u^T (\bm K_\sigma + \bm K_{\ttsigmad})\qcm \\
\bm x\idx{th} &= \bm s_{\pd{\sigma}{T} E^2}\qpt
\end{alignat}
\end{subequations}
The results are collected in Table~\ref{tab:model}. The sensitivities are both positive and negative and their absolute values vary substantially. Comparing the absolute value of the sensitivity to $p_1$ and $p_2$, respectively, indicates that the \ac{qoi} is much more sensitive to changes in $p_1$ compared to $p_2$. However, the comparison of the derivatives should take the absolute value of the parameters into account. This can be done for example by using a first order Taylor series approximation of the \ac{qoi}'s dependence on the parameters, 
\begin{equation*}
G\idx{Joule}(p_j) \approx G\idx{Joule}(p_{0,j}) + \dd{G\idx{Joule}}{p_j}(p_{0,j}) \Delta p_j,\quad j = 1,...,5\qcm
\end{equation*}
where $\Delta p_j = p_j - p_{0,j}$ is the perturbation of the $j$-th parameter. The Taylor series can be used to estimate the relative change of a \ac{qoi} that occurs for an increase of a parameter $\Delta p_j = 1\% p_{0,j}$, i.e.
\begin{equation}
 \Delta G_{1\%, j} := \frac{\Delta G\idx{Joule}}{G\idx{Joule}(p_{0,j}} \approx  \dd{G\idx{Joule}}{p_j}(p_{0,j})\frac{1\% p_{0,j}}{G\idx{Joule}(p_{0,j})}\qcm \label{eq:norm}
\end{equation}
with $\Delta G\idx{Joule} ={G\idx{Joule}(p_{0,j} + \Delta p_j) - G\idx{Joule}(p_{0,j})}$. 
The normalized sensitivities are brought together in Table~\ref{tab:model2}. 
Comparing the normalized value of the derivatives to $p_1$ and $p_2$ now shows that the \ac{qoi} actually is much more sensitive towards {relatively} small changes in $p_2$ compared to $p_1$ (see Fig.~\ref{fig:sweep_wloss_normalized_sigma0} and Fig.~\ref{fig:sweep_wloss_normalized_E1}). 
\begin{table}[tb]
\caption{Sensitivities of the Joule heat with respect to the parameters $p_1$ to $p_5$ of the nonlinear \ac{fgm} conductivity defined by \eqref{Conductivity}.}
\centering
\begin{tabular}{ccc}
\toprule
 \textbf{Parameter } &\textbf{Value} &  \textbf{Derivative} \\
 \textbf{$\bm{p_j}$} &  $\bm{p_{\bm{0,j}}}$&  $\dd{G}{p_j}(\bm p_0)$\\
\midrule
 $p_1$ &  1.0e-10\,S/m & 2.76e+11\,kJ/(S/m) \\
 $p_2$ &  0.70\,kV/mm  &  -6.28e-4\,kJ/(kV/mm)\\
 $p_3$ &  2.4kV/mm &  2.38e-8\,kJ/(kV/mm) \\
  $p_4$ &  1900 &  1.65e+2\,kJ \\
  $p_5$ &  3700 &  1.01e+2\,kJ \\
\bottomrule
\end{tabular}
\label{tab:model}
\end{table}
\begin{table}[tbh]
\caption{Normalized sensitivities according to \eqref{eq:norm} of the Joule heat with respect to the parameters $p_1$ to $p_5$ of the nonlinear \ac{fgm} conductivity defined by \eqref{Conductivity}.}
\centering
\begin{tabular}{ccc}
\toprule
 \textbf{Parameter} &  \textbf{Normalized sensitivity}\\
 \textbf{$\bm{p_j}$} &  \textbf{$\bm{\Delta G_{1\%,j}}$}\\
\midrule
 $p_1$ & $0.780\,\%$ \\
 $p_2$ &  $-12.4\,\%$\\
 $p_3$ &   1.61e-3\,\%\\
  $p_4$ &  $0.870\,\%$ \\
  $p_5$ &  $1.06\,\%$ \\
\bottomrule
\end{tabular}
\label{tab:model2}
\end{table}
Finally, the adjoint formulation \eqref{eq:eqs:adj} and \eqref{eq:thc:adj} is validated by comparing the sensitivities of the Joule heat obtained by the \ac{am} to two reference solutions. The first reference solution is obtained by the commercial simulation software COMSOL Multiphysics$^\circledR$ and the second reference solution is computed using the \ac{dsm} which is also implemented in \textit{Pyrit}. Fig.~\ref{fig:validation_sensitivities} shows that the results agree for all parameters. Hence, the method is successfully validated. 

Figure~\ref{fig:convergence} investigates the convergence behavior of the \ac{am}. Figure~\ref{fig:relerr_wloss_mesh} shows that the relative error, $\epsilon\idx{rel}$, of the sensitivity $\dd G {p_2}(\bm p_0)$ of the Joule losses to $p_2$ converges quadratically with respect to the maximum edge length inside the \ac{2d} triangular mesh. In order to achieve a relative error below one 1\%, mesh consisting of 24722 nodes and 52544 elements is selected for all further computations. 

Figure~\ref{fig:relerr_wloss_dtmax} shows the relative error, $\epsilon\idx{rel}$, of the sensitivity $\dd G {p_1}(\bm p_0)$ of the Joule losses to $p_2$ with respect to the time step size. 
{As expected for the implicit Euler method, the relative error converges quadratically with respect to the time step size.} With a maximum step size of {$\Delta t\idx{max} = 0.56\,\+{ms}$}, a relative error below $0.1\,\%$ can be achieved. It furthermore shows the convergence of the sensitivity with respect to the maximum thermal time step, $\Delta t\idx{th,max}$, while the maximum electric time step is fixed at {$\Delta t\idx{el,max} = 0.56\,\+{ms}$}. The thermal time step can be chosen approximately 5.4 times larger than the electric time step, demonstrating the benefits of a multi-rate time-integration approach. The \ac{am} is, thus, successfully validated, which is an important step towards the optimization of \acp{fgm} in \ac{hvdc} cable joints.
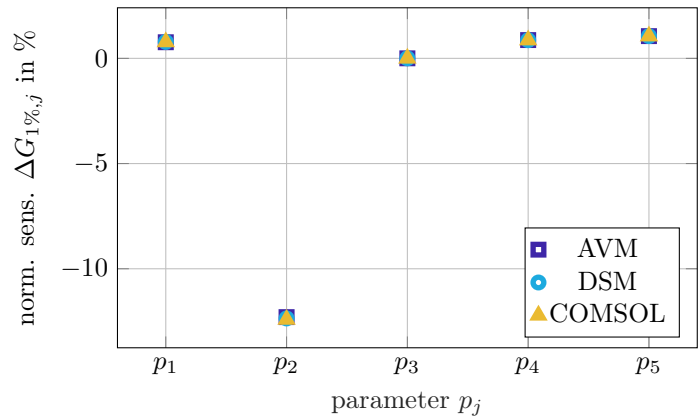
\begin{figure}[tbh]	
	\centering
	\setlength{\fwidth}{0.45\columnwidth}
	\input{figs/validation_sensitivities.tex}\label{fig:validation_sensitivities}
	\caption{Comparison of the sensitivities of the Joule heat computed by the \ac{am} and \ac{dsm}, respectively.}
	\label{fig:validation_sensitivities}
\end{figure}
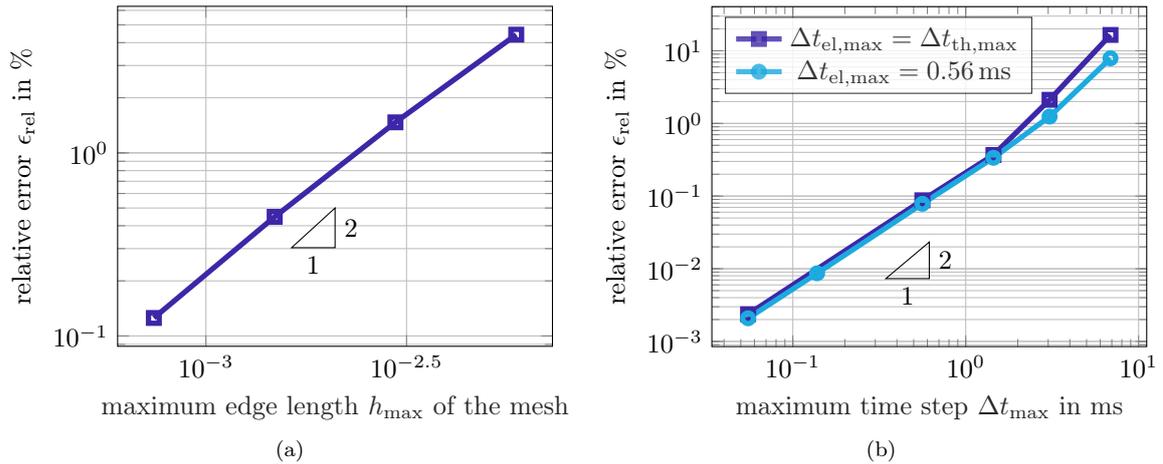
\begin{figure}[tb]
	\setlength{\fwidth}{0.45\columnwidth}	
	\centering
	\subfloat[]{
		\label{fig:relerr_wloss_mesh}\input{figs/relerr_wloss_nnodes.tex}}
		\subfloat[]{
		\label{fig:relerr_wloss_dtmax}\input{figs/relerr_wloss_dtmax.tex}\label{fig:relerr_wloss_dtmax}}	
		\caption{Convergence of the sensitivity $\dd G {p_2}$ of the Joule heat to $p_2$ with respect to (a) the mesh size and (b) the time step size.}\label{fig:convergence}
\end{figure}

%% file: figs/switching_impulse.tex
%
%

\definecolor{mycolor1}{rgb}{0.24220,0.15040,0.66030}%

\begin{tikzpicture}
    \begin{axis}[
width=1.1\fwidth,
height=0.65\fwidth,
at={(0\fwidth,0\fwidth)},
scale only axis,
xlabel style={font=\color{white!15!black}},
yminorticks=true,
axis background/.style={fill=white},
xmajorgrids,
ymajorgrids,
yminorgrids,
    xlabel={time $t$ in ms},
    ylabel={voltage $U_\+{switch}$ in kV}]
        \addplot+[color=mycolor1, line width=2.0pt, mark = none] table[x=x, y=y, x expr = \thisrowno{0} * 1e3, y expr = \thisrowno{1} * 1e-3 + 320] {./figs/switching_impulse.dat};
    \end{axis}
\end{tikzpicture}


%% file: figs/switching_Ez_z_t.tex
%
%

\definecolor{mycolor1}{rgb}{0.24220,0.15040,0.66030}%
\definecolor{mycolor2}{rgb}{0.91840,0.73080,0.18900}%
\definecolor{mycolor3}{rgb}{0.10850,0.66690,0.87340}%
\definecolor{mycolor1}{rgb}{0.242200000000000 ,  0.150400000000000 ,  0.660300000000000}%
\definecolor{mycolor2}{rgb}{0.108500000000000 ,  0.666900000000000  , 0.873400000000000}%
\definecolor{mycolor3}{rgb}{0.280900000000000  , 0.796400000000000   ,0.526600000000000}%
\definecolor{mycolor4}{rgb}{  0.918400000000000 ,  0.730800000000000  , 0.189000000000000}%

\begin{tikzpicture}
    \begin{axis}[
width=1.1\fwidth,
height=0.65\fwidth,
at={(0\fwidth,0\fwidth)},
scale only axis,
xlabel style={font=\color{white!15!black}},
yminorticks=true,
axis background/.style={fill=white},
xmajorgrids,
ymajorgrids,
yminorgrids,
    xlabel={length $z$ in m},
    ylabel={electric field $E_z$ in kV/mm},
    legend style={at={(0.2,0.01)},anchor=south west}]
        \addplot+[color=mycolor1, line width=2.0pt, mark = none] table[x=x, y=y, x expr = \thisrowno{0} * 1, y expr = \thisrowno{1} * 1e-6] {./figs/data_switching_impulse_pyrit/pyrit_Ez_z_t=0.dat}; \addlegendentry{$t = 0.0\,$ms\phantom{$.357$}};
        \addplot+[color=mycolor2, line width=2.0pt, mark = none] table[x=x, y=y, x expr = \thisrowno{0} * 1, y expr = \thisrowno{1} * 1e-6] {./figs/data_switching_impulse_pyrit/pyrit_Ez_z_t=357.dat}; \addlegendentry{$t = 0.36\,$ms};
        \addplot+[color=mycolor3, line width=2.0pt, mark = none] table[x=x, y=y, x expr = \thisrowno{0} * 1, y expr = \thisrowno{1} * 1e-6] {./figs/data_switching_impulse_pyrit/pyrit_Ez_z_t=2500.dat}; \addlegendentry{$t = 2.5\,$ms\phantom{$37$}};
       \addplot+[color=mycolor4, line width=2.0pt, mark = none] table[x=x, y=y, x expr = \thisrowno{0} * 1, y expr = \thisrowno{1} * 1e-6] {./figs/data_switching_impulse_pyrit/pyrit_Ez_z_t=30000.dat}; \addlegendentry{$t = 30\,$ms\phantom{$.57$}}; 
    \end{axis}
\end{tikzpicture}


%% file: figs/sweep_wloss_sigma0.tex
%
%

\definecolor{mycolor1}{rgb}{0.24220,0.15040,0.66030}%

\begin{tikzpicture}
    \begin{axis}[
width=0.75*1.1\fwidth,
height=0.65\fwidth,
at={(0\fwidth,0\fwidth)},
scale only axis,
xlabel style={font=\color{white!15!black}},
xmode=log,
yminorticks=true,
axis background/.style={fill=white},
xmajorgrids,
ymajorgrids,
xlabel={parameter $p_1$ in S/m},
ylabel={Joule heat $G\idx{Joule}$ in J}]
        \addplot+[color=mycolor1, line width=2.0pt, mark = o] table[x=x, y=y, y expr = \thisrowno{1}*2 ] {./figs/data_switching_impulse_pyrit/joule_heat_sweep_sigma0.dat};
    \end{axis}
\end{tikzpicture}


%% file: figs/sweep_wloss_E1.tex
%
%

\definecolor{mycolor1}{rgb}{0.24220,0.15040,0.66030}%

\begin{tikzpicture}
    \begin{axis}[
width=0.75*1.1\fwidth,
height=0.65\fwidth,
at={(0\fwidth,0\fwidth)},
scale only axis,
xlabel style={font=\color{white!15!black}},
yminorticks=true,
axis background/.style={fill=white},
xmajorgrids,
ymajorgrids,
xlabel={parameter $p_2$ in kV/mm},
ylabel={Joule heat $G\idx{Joule}$ in J}]
        \addplot+[color=mycolor1, line width=2.0pt, mark = o] table[x=x, y=y, y expr = \thisrowno{1}*2] {./figs/data_switching_impulse_pyrit/joule_heat_sweep_E1.dat};
    \end{axis}
\end{tikzpicture}


%% file: figs/sweep_wloss_sigma0_normalized.tex
%
%

\definecolor{mycolor1}{rgb}{0.24220,0.15040,0.66030}%

\begin{tikzpicture}
    \begin{axis}[
width=0.75*1.1\fwidth,
height=0.65\fwidth,
at={(0\fwidth,0\fwidth)},
scale only axis,
ytick = {0.99,1,1.01},
xtick = {0.99,1,1.01},
ymin=0.988,
xlabel style={font=\color{white!15!black}},
yminorticks=true,
axis background/.style={fill=white},
xmajorgrids,
ymajorgrids,
xlabel={normalized parameter $p_1 / p_{1,0}$},
ylabel={normalized QoI $\frac{G\idx{Joule}(p_1) }{ G\idx{Joule}(p_{1,0})}$}]
        \addplot+[color=mycolor1, line width=2.0pt, mark = o] table[x=x, y=y, x expr = \thisrowno{0} ] {./figs/data_switching_impulse_pyrit/sweep_wloss_sigma0_normalized.dat};
          \addplot+[color=red, dashed, line width=1.5pt, mark = none] table[x=x, y=y, x expr = \thisrowno{0} ] {./figs/data_switching_impulse_pyrit/slope_sigma0.dat};
    \end{axis}
\end{tikzpicture}


%% file: figs/sweep_wloss_E1_normalized.tex
%
%

\definecolor{mycolor1}{rgb}{0.24220,0.15040,0.66030}%

\begin{tikzpicture}
    \begin{axis}[
width=0.75*1.1\fwidth,
height=0.65\fwidth,
at={(0\fwidth,0\fwidth)},
scale only axis,
xtick = {0.99,1,1.01},
xlabel style={font=\color{white!15!black}},
yminorticks=true,
axis background/.style={fill=white},
xmajorgrids,
ymajorgrids,
xlabel={normalized parameter $p_2 / p_{2,0}$},
ylabel={normalized QoI $\frac{G(p_2) }{ G(p_{2,0})}$}]
        \addplot+[color=mycolor1, line width=2.0pt, mark = o] table[x=x, y=y, x expr = \thisrowno{0} ] {./figs/data_switching_impulse_pyrit/sweep_wloss_E1_normalized.dat};
         \addplot+[color=red, dashed, line width=1.5pt, mark = none] table[x=x, y=y, x expr = \thisrowno{0} ] {./figs/data_switching_impulse_pyrit/slope_E1.dat};
    \end{axis}
\end{tikzpicture}


%% file: figs/validation_sensitivities.tex
%
%

\definecolor{mycolor1}{rgb}{0.24220,0.15040,0.66030}%
\definecolor{mycolor1}{rgb}{0.242200000000000 ,  0.150400000000000 ,  0.660300000000000}%
\definecolor{mycolor2}{rgb}{0.108500000000000 ,  0.666900000000000  , 0.873400000000000}%
\definecolor{mycolor3}{rgb}{0.280900000000000  , 0.796400000000000   ,0.526600000000000}%
\definecolor{mycolor4}{rgb}{  0.918400000000000 ,  0.730800000000000  , 0.189000000000000}%

\begin{tikzpicture}
    \begin{axis}[
width=1.1\fwidth,
height=0.65\fwidth,
at={(0\fwidth,0\fwidth)},
scale only axis,
xlabel style={font=\color{white!15!black}},
xtick={1,2,3,4,5, 6},
xticklabels={$p_1$,$p_2$,$p_3$,$p_4$,$p_5$},
yminorticks=true,
axis background/.style={fill=white},
xmajorgrids,
ymajorgrids,
yminorgrids,
    xlabel={parameter $p_j$},
    ylabel={norm. sens. $\Delta G_{1\%, j}$ in $\%$},
    legend style={legend pos = south east}]
        \addplot+[color=mycolor1, only marks, line width=2.0pt, mark = square] table[x=x, y=y, y expr = \thisrowno{1} * 1e2] {./figs/data_switching_impulse_pyrit/sensitivities_normalized_am.dat}; \addlegendentry{AVM};
        \addplot+[color=mycolor2, only marks, line width=2.0pt, mark = o] table[x=x, y=y, y expr = \thisrowno{1} * 1e2] {./figs/data_switching_impulse_pyrit/sensitivities_normalized_dsm.dat}; \addlegendentry{DSM};
        \addplot+[color=mycolor4, only marks, line width=2.0pt, mark = triangle] table[x=x, y=y, y expr = \thisrowno{1} * 1e2] {./figs/data_switching_impulse_pyrit/comsol_normalized.dat}; \addlegendentry{COMSOL};
    \end{axis}
\end{tikzpicture}


%% file: figs/relerr_wloss_nnodes.tex
%
%

\definecolor{mycolor1}{rgb}{0.24220,0.15040,0.66030}%
\definecolor{mycolor1}{rgb}{0.242200000000000 ,  0.150400000000000 ,  0.660300000000000}%
\definecolor{mycolor2}{rgb}{0.108500000000000 ,  0.666900000000000  , 0.873400000000000}%
\definecolor{mycolor3}{rgb}{0.280900000000000  , 0.796400000000000   ,0.526600000000000}%
\definecolor{mycolor4}{rgb}{  0.918400000000000 ,  0.730800000000000  , 0.189000000000000}%

\begin{tikzpicture}
\newcommand{\logLogSlopeTriangle}[5]
{

    \pgfplotsextra
    {
        \pgfkeysgetvalue{/pgfplots/xmin}{\xmin}
        \pgfkeysgetvalue{/pgfplots/xmax}{\xmax}
        \pgfkeysgetvalue{/pgfplots/ymin}{\ymin}
        \pgfkeysgetvalue{/pgfplots/ymax}{\ymax}

        \pgfmathsetmacro{\xArel}{#1}
        \pgfmathsetmacro{\yArel}{#3}
        \pgfmathsetmacro{\xBrel}{#1-#2}
        \pgfmathsetmacro{\yBrel}{\yArel}
        \pgfmathsetmacro{\xCrel}{\xArel}

        \pgfmathsetmacro{\lnxB}{\xmin*(1-(#1-#2))+\xmax*(#1-#2)} 
        \pgfmathsetmacro{\lnxA}{\xmin*(1-#1)+\xmax*#1} 
        \pgfmathsetmacro{\lnyA}{\ymin*(1-#3)+\ymax*#3} 
        \pgfmathsetmacro{\lnyC}{\lnyA+#4*(\lnxA-\lnxB)}
        \pgfmathsetmacro{\yCrel}{\lnyC-\ymin)/(\ymax-\ymin)} 

        \coordinate (A) at (rel axis cs:\xArel,\yArel);
        \coordinate (B) at (rel axis cs:\xBrel,\yBrel);
        \coordinate (C) at (rel axis cs:\xCrel,\yCrel);

        \draw[#5]   (A)-- node[pos=0.5,anchor=north] {1}
                    (B)-- 
                    (C)-- node[pos=0.5,anchor=west] {#4}
                    cycle;
    }
}

    \begin{axis}[
width=0.75*1.1\fwidth,
height=0.65\fwidth,
at={(0\fwidth,0\fwidth)},
scale only axis,
xlabel style={font=\color{white!15!black}},
ymode=log,
xmode=log,
yminorticks=true,
axis background/.style={fill=white},
xmajorgrids,
ymajorgrids,
yminorgrids,
    xlabel={maximum edge length $h\idx{max}$ of the mesh},
    ylabel={relative error $\epsilon\idx{rel}$ in $\%$},
    legend style = {legend pos = south east, opacity=0.8}]
        \addplot+[color=mycolor1, line width=2.0pt, mark = square] table[x=x, y=y, x expr = \thisrowno{0}, y expr = \thisrowno{1} * 1e3 ] {./figs/data_switching_impulse_pyrit/convergence_mesh_relerr_dwlossdE1.dat}; 
      \logLogSlopeTriangle{0.5}{0.1}{0.29}{2}{black};
    \end{axis}
\end{tikzpicture}


%% file: figs/relerr_wloss_dtmax.tex
%
%

\definecolor{mycolor1}{rgb}{0.24220,0.15040,0.66030}%
\definecolor{mycolor1}{rgb}{0.242200000000000 ,  0.150400000000000 ,  0.660300000000000}%
\definecolor{mycolor2}{rgb}{0.108500000000000 ,  0.666900000000000  , 0.873400000000000}%
\definecolor{mycolor3}{rgb}{0.280900000000000  , 0.796400000000000   ,0.526600000000000}%
\definecolor{mycolor4}{rgb}{  0.918400000000000 ,  0.730800000000000  , 0.189000000000000}%

\begin{tikzpicture}
\newcommand{\logLogSlopeTriangle}[5]
{

    \pgfplotsextra
    {
        \pgfkeysgetvalue{/pgfplots/xmin}{\xmin}
        \pgfkeysgetvalue{/pgfplots/xmax}{\xmax}
        \pgfkeysgetvalue{/pgfplots/ymin}{\ymin}
        \pgfkeysgetvalue{/pgfplots/ymax}{\ymax}

        \pgfmathsetmacro{\xArel}{#1}
        \pgfmathsetmacro{\yArel}{#3}
        \pgfmathsetmacro{\xBrel}{#1-#2}
        \pgfmathsetmacro{\yBrel}{\yArel}
        \pgfmathsetmacro{\xCrel}{\xArel}

        \pgfmathsetmacro{\lnxB}{\xmin*(1-(#1-#2))+\xmax*(#1-#2)} 
        \pgfmathsetmacro{\lnxA}{\xmin*(1-#1)+\xmax*#1} 
        \pgfmathsetmacro{\lnyA}{\ymin*(1-#3)+\ymax*#3} 
        \pgfmathsetmacro{\lnyC}{\lnyA+#4*(\lnxA-\lnxB)}
        \pgfmathsetmacro{\yCrel}{\lnyC-\ymin)/(\ymax-\ymin)} 

        \coordinate (A) at (rel axis cs:\xArel,\yArel);
        \coordinate (B) at (rel axis cs:\xBrel,\yBrel);
        \coordinate (C) at (rel axis cs:\xCrel,\yCrel);

        \draw[#5]   (A)-- node[pos=0.5,anchor=north] {1}
                    (B)-- 
                    (C)-- node[pos=0.5,anchor=west] {#4}
                    cycle;
    }
}

    \begin{axis}[
width=0.75*1.1\fwidth,
height=0.65\fwidth,
at={(0\fwidth,0\fwidth)},
scale only axis,
xlabel style={font=\color{white!15!black}},
ymode=log,
xmode=log,
yminorticks=true,
axis background/.style={fill=white},
xmajorgrids,
ymajorgrids,
yminorgrids,
    xlabel={maximum time step $\Delta t\idx{max}$ in ms},
    ylabel={relative error $\epsilon\idx{rel}$ in $\%$},
    legend style = {legend pos = north west, opacity=0.8}]
        \addplot+[color=mycolor1, line width=2.0pt, mark = square] table[x=x, y=y, x expr = \thisrowno{0} * 1e3, y expr = \thisrowno{1} * 1e2 ] {./figs/data_switching_impulse_pyrit/relerr_dwlossdE1_dtmax_gen_ntref_750.dat}; \addlegendentry{$\Delta t\idx{el,max}=\Delta t\idx{th,max}$};
         \addplot+[color=mycolor2, line width=2.0pt, mark = o] table[x=x, y=y, x expr = \thisrowno{0} * 1e3, y expr = \thisrowno{1} * 1e2 ] {./figs/data_switching_impulse_pyrit/relerr_dwlossdE1_dtmax_gen_ntref_500.dat}; \addlegendentry{$\Delta t\idx{el,max}=0.56\,$ms};
      \logLogSlopeTriangle{0.5}{0.1}{0.2}{2}{black};
    \end{axis}
\end{tikzpicture}


%% file: conclusion.tex
The \acl{am} is an efficient approach for computing sensitivities of quantities of interest with respect to a large number of the design parameters. In this work, the \acl{am} is formulated for coupled transient electrothermal problems with nonlinear media and implemented in the \acl{fe} framework \textit{Pyrit}. The method is applied to the numerically challenging example of a 320\,kV cable joint specimen under switching operation. The computed sensitivities are compared to results obtained by the \acl{dsm} and the method is successfully validated. The convergence of the \acl{am} is discussed and the benefits of a multi-rate time-integration approach are demonstrated. This is an important step towards efficient and systematic design and optimization of \acl{hvdc} cable joints. 